\documentclass[sigconf]{acmart}

\usepackage{booktabs} % For formal tables

\setcopyright{acmcopyright}
\acmDOI{http://dx.doi.org/10.1145/3063386.3063765}

\acmISBN{123-4567-24-567/08/06}

\acmConference[SCOPE 2017]{The 2nd Workshop on Science of Smart City Operations and Platforms Engineering}{21 April 2017}{Pittsburgh, PA USA} 
\copyrightyear{2017}

\acmPrice{15.00}

\begin{document}
\title{A Colonel Blotto Game for Interdependence-Aware Cyber-Physical Systems Security in Smart Cities}

\author{ Aidin Ferdowsi, Walid Saad}
\affiliation{%
  \institution{Wireless@VT, Bradley Department of Electrical and Computer Engineering}
  \city{Blacksburg} 
  \state{VA} 
  \postcode{24061}
  \country{USA}
}
\email{{aidin,walids}@vt.edu}

\author{Behrouz Maham}
\affiliation{%
  \institution{School of Engineering, Nazarbayev University}
  \city{Astana} 
  \country{Kazakhstan}
}
\email{behrouz.maham@nu.edu.kz}

\author{Narayan B. Mandayam}
\affiliation{
  \institution{WINLAB, Rutgers University}
  \city{North Brunswick}
  \state{NJ}
  \postcode{08902} 
  \country{USA}
}
\email{narayan@winlab.rutgers.edu}

\thanks{This research was supported by the U.S. National Science Foundation under Grants ACI-1541105 and ACI-1541069.} 
\renewcommand{\shortauthors}{A. Ferdowsi et al.}
\renewcommand{\shorttitle}{A Colonel Blotto Game for Interdependent CPS Security}

\begin{abstract}
Smart cities must integrate a number of interdependent cyber-physical systems that operate in a coordinated manner to improve the well-being of the city's residents. A cyber-physical system (CPS) is a system of computational elements controlling physical entities. Large-scale CPSs are more vulnerable to attacks due to the cyber-physical interdependencies that can lead to cascading failures which can have a significant detrimental effect on a city. In this paper, a novel approach is proposed for analyzing the problem of allocating security resources, such as firewalls and anti-malware, over the various cyber components of an interdependent CPS to protect the system against imminent attacks. The problem is formulated as a Colonel Blotto game in which the attacker seeks to allocate its resources to compromise the CPS, while the defender chooses how to distribute its resources to defend against potential attacks. To evaluate the effects of defense and attack, various CPS factors are considered including human-CPS interactions as well as physical and topological characteristics of a CPS such as flow and capacity of interconnections and minimum path algorithms. Results show that, for the case in which the attacker is not aware of the CPS interdependencies, the defender can have a higher payoff, compared to the case in which the attacker has complete information. The results also show that, in the case of more symmetric nodes, due to interdependencies, the defender achieves its highest payoff at the equilibrium compared to the case with independent, asymmetric nodes. 
\end{abstract}



% We no longer use \terms command


\maketitle
\vspace*{-0.2cm}\section{Introduction}
Smart cities encompass a multitude of cyber-physical systems (CPSs), such as smart grids and smart transportation systems, that are critical to enhancing municipal services and reducing living costs for residents \cite{ ref_2, ref_3, ref_4}. One key feature of CPSs is the strong synergy between their cyber and physical functions. For instance, the compromise of a communication node in a transportation system can have a direct physical impact on the involved vehicles. Thus, modeling and analyzing failures across an interdependent CPS in a smart city is a problem of critical and global importance \cite{ref_6} and \cite{ref_15}. Protecting CPSs faces numerous challenges such as interdependence between cyber and physical failures, modeling the key functional characteristics of a CPS, and appropriately responding to CPS attacks.

Recently, the security of CPSs has received significant attention in the existing literature \cite{ref_3, ref_4, ref_5, ref_6, ref_7, ref_8, ref_9, ref_10, ref_11, ref_12, ref_15,ref_16}. The first body of work in \cite{ref_8, ref_9, ref_10} focused on the game-theoretic analysis of CPS security. The authors in \cite{ref_8} formulated a three-stage Colonel Blotto game with hierarchical information structure, in which two players fight against a common adversary and have presented this game as an applicable model to cyber vulnerability in power systems. In \cite{ref_9}, the authors studied a resilient control problem, in which control packets transmitted over a network are corrupted by a human adversary. A Stackelberg game is proposed to stabilize the control system despite the attack. In \cite{ref_10}, the equilibrium of a heterogeneous Blotto game is investigated for asymmetric battlefields having different values. 

Beyond game-theoretic analysis, recent works such as \cite{ref_7} have also analyzed the control-theoretic aspect of CPS security.  In \cite{ref_16}, the authors proposed general models for defenders and probable attacks to a CPS. By using graph theory, the components of CPSs were modeled with the vertices being physical or cyber elements and the graph edges being interconnections.  In \cite{ref_3} and \cite{ref_15}, the authors studied bad data injection attacks in a smart grid and analyzed the impact of such attacks on the physical side of the grid. The work in \cite{ref_11} presented a model-based design methodology with the focus on physical properties by the case study of a tunneling ball device. In \cite{ref_12}, a general CPS model with cascading failures is studied. However, despite being interesting, these existing works \cite{ref_7, ref_8, ref_9, ref_10, ref_11, ref_12,ref_16} do not explicitly consider the interdependencies of cyber and physical elements during security resource allocation. Moreover, this body of work has not considered the topological characteristics when analyzing security threats on CPSs. Moreover, existing game-theoretic studies such as in \cite{ref_10} did not take into account the effects of nodes on each other. Also, the game analysis in \cite{ref_10} ignores the physical characteristics of the CPS nodes.

The main contribution of this paper is to introduce a novel approach, based on the Colonel Blotto game \cite{ref_13}, to allocate security resources across the interdependent elements of a CPS within a smart city. In particular, we use graph theory to model the elements of a CPS as connected nodes. In this model, we explicitly take into account human-CPS interaction, the load carried over a CPS,  as well as the flow and capacities of interconnections. Then, we formulate the security resource allocation problem as a Colonel Blotto game between a defender and an attacker. In this game, the attacker seeks to compromise most of the physical nodes by allocating its attack resources over the cyber nodes to maximize its payoff while the defender aims to minimize the number of compromised nodes to minimize its loss. For this game, we investigate the mixed-strategy Nash equilibrium (NE) and the payoff for each player for two cases: a case in which the attacker is aware of interdependencies and a case in which it is not aware of those interdependencies. Then, we prove that, under a symmetric game, the defender's payoff is higher at the NE compared to the asymmetric game. Simulation results assess the various properties of the game and the impact of attack and defense on the CPS, under different scenarios.

The rest of this paper is organized as follows. Section II presents the studied CPS model. In Section III, the Colonel Blotto formulation is presented. In Section IV, simulation results and a case study are presented while Section V draws some conclusions.

\section{System Model and Problem Formulation}\label{sec:Model}
\subsection{Cyber-Physical System Model}
To model a general CPS, we use four common features that are found across a wide range of CPSs \cite{ref_4}, \cite{ref_11}, and \cite{ref_14}: 
\begin{itemize}
	\item\emph{Load:} The load that is delivered from a point in a physical network to another point such as power in smart grids or vehicles in a transportation system.
	\item\emph{Flow:} The flow is the rate of load which is transmitted over interconnections. Examples include the number of cars passing a street at, the rate of food entity transmission from different devices in an automated factory, or the power delivery rate in power systems.
	\item\emph{Capacity:} The maximum capacity of flow is a key CPS parameter that is determined by physical limitations within a CPS such as the maximum speed of cars or the maximum power transmission in power systems.
	\item\emph{Human Interaction:} Humans influence a CPS by being a part of its physical system or being served by it within a smart city. Human interactions, are, thus, a key feature needed to assess the importance of nodes in a CPS.
\end{itemize}
Given these features, a CPS can be modeled as a network of computing and cyber elements which are controlling a network of physical system. Hereinafter, we use term ``node" to describe both cyber and physical elements. In addition, each physical system in a CPS consists of three main types of elements: one  reference node as well as a number of main nodes and ordinary nodes. Practical examples include slack, PV, and PQ buses in power systems \cite{ref_14} and central square or subway station and the surrounding stations in a \emph{concentric city} design such as in the Berlin subway system. In our model, we suppose that each physical node is controlled by one cyber node. Then, the graphical representations of physical and cyber systems will be analogous. 
\subsection{Value of each node in a CPS}
Each cyber node has a value depending on its level of interaction with the humans in the CPS. Meanwhile, each physical element can be monitored or operated by actual individuals. Finding practical values that capture the importance of each CPS node will facilitate the analysis. We consider the human-CPS interaction as an appropriate feature to determine the value of any given CPS node. Therefore, to characterize the interactions between humans and the CPS, we consider the population level. In particular, the presence of a larger population at a physical node results in more interactions between individuals and the CPS. Thus, we can assume that the nodes which are located in the \emph{downtown area of a city} have more human interaction than others. Here, we let $h_i$ be the fraction of human interaction at node $i$ out of the total human interactions at all of the nodes in the physical system. In a concentric design, $h_i$ will be a function of the distance from the most populated points of the city. Given this model, next, we characterize how the failure of a particular node will impact the entire system, and then we use that characterization to define a parameter that represents the interdependencies of the CPS nodes. Such a parameter will then capture the dependence between cyber and physical elements.

For the \emph{physical system} of our CPS, we consider a set $\mathcal{N}$ of $n$ physical nodes. Here, we assume that the total input flow to each physical node is equal to the output flow of this node, except in the marginal nodes of the system. Moreover, since a physical system can extend over a vast area and physical nodes are relatively distant from one another, we assume that a failure in a physical node can only impact the flow in nearby nodes. In the considered CPS, the total flow of the system should be maintained as constant following any failure or attack. Also, the direction of the flow will be one-sided and starts from the reference node towards ordinary nodes.

Next, we present a method to analyze the effect of the failure of a physical node on the entire physical system. We introduce a matrix an $n \times n$ matrix $\boldsymbol{F}$ to represent the flow of the edges between each pair of nodes. Each element , $f_{ij}$, of $\boldsymbol{F}$ is positive if, for two connected nodes, node $i$ is closer than node $j$ to the reference node and it is zero otherwise. Therefore, the positive elements in row $j$ are the nodes which are being supplied with the flow from node $i$ and the positive elements in column $i$ are the nodes which are supplying node $i$. After a failure in node $i$ both column $i$ and row $i$ will have only zero elements. Then, we let $\boldsymbol{F}_i$ be a matrix which is identical to $\boldsymbol{F}$ but where its column $i$ and row $i$ are zero.

To start analyzing the effect of changing column $i$ and row $i$ to zero (which implies that there is no input or output flow in node $i$ of the failed system), first, we define two sets of nodes: first-order distant nodes which are directly connected to node $i$, $\mathcal{N}_1^i=\{j \in \mathcal{N} |f_{ij}>0 \cup f_{ji}>0 \}$ and second-order distant nodes which are connected to the first-order distance nodes of $i$, $\mathcal{N}_2^i=\{ k \in \mathcal{N} |\forall j\in \mathcal{N}_1^i , f_{kj}>0 \cup f_{jk}>0 \}$. To characterize the effect of a failure at node $i$, we start with the set of nodes in $\mathcal{N}_1^i$ which are supplied by node $i$ which is denoted by $\mathcal{M}^i=\{j \in \mathcal{N} |f_{ij}>0\}$ and, thus $\mathcal{M}^i \subseteq\mathcal{N}_1^i$. The analysis can proceeds as follows:

\begin{enumerate}
	\item We choose nodes in  $\mathcal{M}^i$ which are not connected to any other nodes in this set or $\{j \in \mathcal{M}^i | \nexists k  \in \mathcal{M}^i , f_{jk}>0\}$.
	\item We find the amount of increase in the flow of whole of the connected edges to node $j$ after a failure in node $i$:
	\begin{equation}\label{eq:1}
	\sum^{n}_{k=1,k\neq i}\delta _{kj}=f_{ij}+\left( \sum^{n}_{k=1,k\neq i}f_{jk}- \sum^{n}_{k=1,k\neq i}f_{kj} \right).
	\end{equation}
	The new flows ($f^{\mathrm{new}}_{kj} , k \neq i$) will be given by:
	\begin{equation}\label{eq:2}
	f^{\mathrm{new}}_{kj}=\begin{cases}
	c_{kj}, &\textrm{ if } f_{kj} > \sum^{n}_{k=1,k\neq i}d_{kj},\\
	f_{kj}+\frac{d_{kj}}{\sum^{n}_{k=1,k\neq i}d_{kj}}\times\\ \sum^{n}_{k=1,k\neq i}\delta _{kj}, & \text{otherwise,}
	\end{cases}
	\end{equation}
	where $c_{kj}$, $f_{kj}$, $d_{kj}$ and $\delta _{kj}$ are, respectively, the capacity, flow, difference between the current flow and capacity, and difference between the flow before and after a failure.
	\item After computing the new flows, the nodes in $\mathcal{M}^i$ are omitted and the same analyze of step 1) is applied to $\mathcal{M}_\textrm{new}^i=\mathcal{M}^i-\{j \in \mathcal{M}^i | \nexists k  \in \mathcal{M}^i , f_{jk}>0\}$. These iterations continue until $\{j \in \mathcal{M}^i | \nexists k  \in \mathcal{M}^i , f_{jk}>0\}=\emptyset$.
	\item The set of nodes such that $\{j \in \mathcal{M}^i | \nexists k  \in \mathcal{M}^i , f_{kj}>0\}$ are choosen and the same approach of step 2) and 3) is applied to this set. This will proceed until $\{j \in \mathcal{M}^i | \nexists k  \in \mathcal{M}^i , f_{kj}>0\}=\emptyset$.
	\item The final step is to choose the set of nodes such that  $\{j \in \mathcal{N}_2^i | \nexists k  \in \mathcal{N}_2^i , f_{kj}>0\}$ and we find new flows for them as in 2) and 3). 
\end{enumerate}

This process continues until $\{j \in \mathcal{N}_2^i | \nexists k  \in \mathcal{N}_2^i , f_{kj}>0\}=\emptyset$. Once the above approach is completed, we can find the total input and output loss of flow , $f^i_j$, at a given physical node $j$ following the failure of node $i$ as follows:
\begin{equation}\label{eq:3}
f^{i}_{j}=\sum_{k=1}^{n}f_{jk}-\sum_{k=1,k\neq i}^{n}f^{\mathrm{new}}_{jk}.
\end{equation}

Now, we can define effect of the failure of node $i$ on node $j$ in the physical system as the fraction of its failure loss to the total flow of node $j$ before the failure as follows:
\begin{equation}\label{eq:4}
e_{ji}=\frac{f^{i}_{j}}{\sum_{k=1}^{n}f_{jk}}=1-\frac{\sum_{k=1,k\neq i}^{n}f^{\mathrm{\mathrm{new}}}_{jk}}{\sum_{k=1}^{n}f_{jk}},
\end{equation}
where $e_{ji}$ is the failure effect of node $i$ on the physical node $j$.

Next, we analyze the effect of the failure of a \emph{cyber node} on the cyber system. We consider that each physical node is being controlled by one cyber node. Unlike the physical system, the distance between the nodes is not a limitation here. However, the failure in one cyber node can affect all of the cyber nodes, since cyber nodes are connected through communication links and exhibit a lower delay than in the case of physical links.

Here, we use the shortest path among the nodes to find a value for each cyber node. Note that due to the one-to-one model, the set of cyber nodes is $\mathcal{N}$ similar to the physical nodes. Thus, we find all pairs of shortest path solutions for the set of nodes $\mathcal{N}$ and $\mathcal{N}-\{i\}$. Then, we find the increase in the summation of all pairs of shortest paths after removing a cyber node from the cyber network. We assume that each pair of the nodes in $\mathcal{N}$ has the minimum path $p_{jk}$ and each pair in $\mathcal{N}-\{i\}$ has the minimum path $p^i_{jk}$. Also, removing a node from the cyber network increases the computational load at each node. If we denote the entire load of computations in the cyber network by $C_L$, then the calculations per each node will be $\frac{C_L}{n}$ where $n$ is the number of all the cyber nodes. We, then define the ratio of the value of each cyber node to that of other nodes in the cyber network based on the failure effect on each:
\begin{equation}\label{eq:5}
t_{ji}=\frac{\sum^{N}_{k=1,k\neq i}p^{i}_{jk}}{\sum^{N}_{k=1}p_{jk}}-1 + t^0.
\end{equation}

In (\ref{eq:5}), we can see that each node has a minimum effect of $t^0$ on any other node due to the increase in the computational load on every node after a failure at a specific node.

As mentioned earlier, in addition to the value of each node, we can assign a value for the interconnection between two nodes. In this context, any failure in each of the cyber or physical nodes will affect both the cyber and physical components of the CPS. Therefore, we define the effect of failure on the whole system using the parameters that we have defined thus far, and we refer to it as the \emph{interdependency parameter} between two cyber nodes $i$ and $j$:
\begin{equation}\label{eq:6}
v_{ji}=\alpha e'_{ji}+ \beta t'_{ji}.
\end{equation}
where  $e'_{ji}$ and $t'_{ji}$ are normalized $e_{ji}$ and $t_{ji}$ respectively and $\alpha+\beta=1$, with both $\alpha$ and $\beta$ being positive constants for normalizing the interconnection parameter.\vspace{-0.05cm}
\subsection{Problem Definition}
CPSs are vulnerable to attacks in both cyber and physical realms. Further, the interdependencies between the physical and cyber elements as captured by (\ref{eq:6}) increase the potential of attacks to the CPS rendering it more arduous to be defended. In our model, we consider an attacker that is distributing attack resources such as malware and trojan horses over the cyber nodes to compromise the CPS and exploit its interdependencies.

While the attacker tries to allocate its restricted destructive resources over the cyber nodes, the owner of the CPS will act as a defender that seeks to optimally allocate its defense resources such as anti-viruses and malware detectors to prevent the attacker from causing a long-lasting failure on the system. Also, due to the varying importance of each CPS node and the intensity of interdependencies among the nodes, analyzing how the attacker and defender will interact over the CPS and allocate their resources is a challenging CPS problem that we study here using the game-theoretic Colonel Blotto framework \cite{ref_13}.

\section{Game Formulation and Solution}\label{sec:Game}
To model the security resource allocation problem in a CPS, we consider the interactions between the defender, referred to as player $D$, and the attacker, referred to as player $A$, using the powerful framework of a Colonel Blotto game \cite{ref_13}. This framework studies the interactions between two generals that seek to allocate limited resources across a number of battlefields. The general that ends up with more resources in a given battlefield will win it. In the classical Blotto game, the winner is the general who wins the most battlefields. Here, we exploit this analogy between the Blotto game and our CPS security resource allocation problem, and we consider that the defender and attacker interact over \emph{the cyber nodes of the CPS, which are hereinafter referred to as battlefields}. Unlike the classical Blotto game in\cite{ref_17}, our game has \emph{asymmetric} values for the battlefields. Moreover, the interdependence between battlefields, due to the CP interconnections, renders our problem significantly different from the classical Blotto models such as in \cite{ref_11}.

The number of battlefields in our game is $n$ which is the number of nodes in the cyber network. The total amount of resources available for the defender and attacker are $R_D \in \mathbb{R_+}$ and $R_A \in \mathbb{R_+}$, respectively with  $R_D\geq R_A$. A particular allocation by player $p \in \{A,D\}$ is defined by the non-negative $n$-dimensional vector $\boldsymbol{r}^p:=[r_1^p,r_2^p,\dots,r_n^p]$ where $r_m^p \ge 0$ is the amount of resources allocated to the $m$-th battlefield by player $p$. Thus, the set of feasible allocations for defender and attacker, $\mathcal{B}^p$, are
\begin{equation}\label{eq:beta1}
\mathcal{B}^p:=\Bigg\{\boldsymbol{r}^p \in \mathbb{R}_+^n\Bigg | \sum_{m=1}^{n}r_m^p = R^p\Bigg\}.
\end{equation}
While a pure strategy for player $p $ would be a deterministic choice between one of the cases in $\mathcal{B}^p$, a mixed strategy for player $p$ is defined as the $n$-variate joint distribution function $F_p : \mathbb{R}_+^n \rightarrow [0, 1]$ with support in $\mathcal{B}^p$. This mixed strategy presents the probability of allocating a fraction of $R_p$ over each of the battlefields. Note that any joint distribution may be broken into a set of univariate marginal distribution functions. To determine the utility function of the players, we will take into account the CPS model and parameters defined in Section II. As mentioned previously, the attacker tries to disrupt the functioning of a CPS by compromising its cyber nodes. Each battlefield has a constant value based on human interaction which we refer to as $h_i$ and there exists $n$ battlefields. Also, in this game each battlefield will have an effect on other battlefields, and hence winning a battlefield leads to gain from other battlefields. However, the attacker has no information about the interconnections. The utility achieved by the defender for protecting each battlefield will be:
\begin{equation}\label{eq:util defender}
u_i^D(r_i^D,r_i^A)=\begin{cases}
g_i, &\textrm{ if } r_i^D > r_i^A ,\\
0, & \text{otherwise,}
\end{cases}
\end{equation}
where $g_i$ is:\vspace*{-0.25cm}
\begin{equation}\label{eq:newval}
g_i=\frac{h_i+\sum_{i=j,j\ne i}^{n}v_{ji}h_j}{\sum_{i=1}^{n}(h_i+\sum_{i=j,j\ne i}^{n}v_{ji}h_j)} .
\end{equation}
For the attacker, as it has no information about the interdependencies, the utility will be:\vspace*{-0.25cm}
\begin{equation}\label{eq:util attacker}
u_i^A(r_i^A,r_i^D)=\begin{cases}
h_i&\textrm{ if } r_i^A > r_i^D, \\
0, & \text{otherwise.}
\end{cases}
\end{equation}
Then, the payoffs for the defender and attacker are given by:\vspace*{-0.25cm}
\begin{equation}\label{payoff_defender}
\mathrm{\pi}_D\left(\boldsymbol{r}^D,\{G_{A,i}\}_{i=1}^n\right)\hspace*{-0.3em}=\hspace*{-0.3em}\sum_{i=1}^ng_iG_{A,i}(r_i^D),
\end{equation}\vspace*{-0.25cm}
\begin{equation}\label{payoff_attacker}
\mathrm{\pi}_A\left(\boldsymbol{r}^A,\{G_{D,i}\}_{i=1}^n\right)\hspace*{-0.3em}=\hspace*{-0.3em}\sum_{i=1}^nh_iG_{D,i}(r_i^A).
\end{equation}
Here, the marginal distribution functions of each battlefield are $G_{p,i}$ where $p$ is the player index and $i$ is the index of the battlefield. These marginal distribution functions capture the probability of allocating a particular fraction of resources over each one of the battlefields. Each player tries to maximize its expected payoff, and therefore solving the Blotto game reduces to finding $G_{p,i}$.

In \cite{ref_13}, a general solution for the continuous Blotto game with asymmetric battlefields was presented. Asymmetric battlefields as in our case imply battlefields that lead to different gains for the player that wins them. It is proved in \cite{ref_13} that, for an attacker and defender with $h_i$ and $g_i$ as the value of battlefields, marginal distribution functions that maximize the payoffs will be:
\begin{equation}\label{attackerMDF}
G_{A,i}(r)=\left(\frac{\frac{g_i}{\lambda _D}-\frac{h_i}{\lambda_ A}}{\frac{g_i}{\lambda _D}}\right)+\frac{r}{\frac{g_i}{\lambda _D}} \: \: \: r \in \left[0,\frac{h_i}{\lambda _A}\right],
\end{equation}
\begin{equation}\label{defenderMDF}
G_{D,i}(r)=\frac{r}{\frac{h_i}{\lambda _A}} \: \: \: r \in \left[0,\frac{h_i}{\lambda _A}\right],
\end{equation}
where $\lambda _p$ is the multiplier on player $p$'s resource expenditure. Also, to find these variables, we must take into account the budget constraint in \eqref{eq:beta1} which yields:
\begin{equation}\label{eq:attackrestrict}
\sum_{i \in \Omega _A} \frac{g_i}{2\lambda_ D}+\sum_{i\notin \Omega _A} \frac{\left(\frac{h_i}{\lambda _A}\right)^2}{2\left(\frac{g_i}{\lambda _D}\right)}=R_A,
\end{equation}
\begin{equation}\label{eq:defenderrestrict}
\sum_{i \in \Omega _A} \frac{\left(\frac{g_i}{\lambda _D}\right)^2}{2\left(\frac{h_i}{\lambda _A}\right)}+\sum_{i\notin \Omega _A}  \frac{h_i}{2\lambda_ A}=R_D,
\end{equation}
where $\Omega _A$ denotes the set of battlefields in which $\frac{h_i}{g_i}>\frac{\lambda _A}{\lambda _D} $. To find $\lambda _A$ and $\lambda _D$, first, we define $\mu \equiv \frac{\lambda _A}{\lambda _D}$ and by then by taking the ration of (\ref{eq:attackrestrict}) to (\ref{eq:defenderrestrict}), we will have:
\begin{equation}\label{eq:main}
\mu ^3 \sum_{i \in \Omega _A} \frac{\left(g_i\right)^2}{h_i}-\mu ^2 \frac{R_D}{R_A}\sum _{i \in \Omega _A}g_i+\mu \sum_{i\notin \Omega _A}h_i-\frac{R_D}{R_A}\sum_{i\notin \Omega _A} \frac{\left(h_i\right)^2}{g_i}=0.
\end{equation}

In \cite{ref_13}, it is shown that for each solution of (\ref{attackerMDF}) and (\ref{defenderMDF}) there exists only one Nash equilibrium. (\ref{eq:main}) does not have a general closed-form solution as it depends on the values of $\frac{h_i}{g_i}$. However, next, we solve this equation for a special case of interest to the CPS problem at hand.
\section{Analysis and Numerical Results for CPS with Interdependencies} \vspace*{-0.2cm}
\subsection{Analytical Results}
First, we will study the solution of (\ref{eq:main}) for a particular case in which the interdependencies only alter the value of the node with the maximum human-CPS interaction. Let $h_m$ and $h_l$ be, respectively, the maximum and minimum human interaction. In the case of maximum $g_i$, we have:
\begin{equation}\label{maxh}
\mathrm{max}\: \{g_i\}=\frac{h_m+h_l}{1+h_l}.
\end{equation}
This happens if $v_{ji}=1$, if $i=m$, and $j=l$ and $0$, otherwise.
This means that failure in the node with the maximum human interaction results in compromising the node with lowest human interaction and has no effects on other nodes. For the case of maximum feasible $g_i$, we will have:
\begin{equation}\label{ratioinmax}
\frac{h_i}{g_i}=\begin{cases}
\frac{h_l(l+h_l)}{h_l+h_m},&\textrm{ if } i=m,\\
1+h_l, & \text{otherwise.}
\end{cases}
\end{equation}

In this case, to solve (\ref{eq:main}), we have three conditions for $\mu$: $\mu\geq 1+h_l$, $\frac{h_l(l+h_l)}{h_l+h_m} \leq \mu< 1+h_l$, and $\mu<\frac{h_l(l+h_l)}{h_l+h_m}$. Next, we consider $\mu \leq 1+h_l$ to analyze the increase in the payoff for the defender if we compare the condition of no information to the condition of complete information of interdependency for attacker.
\vspace*{-0.1cm}
\begin{theorem}\label{theorem:1}
	At equilibrium, compared to the case in which it has complete information, the expected payoff for the defender increases if: a) the attacker has no information about the interdependencies, b) $g_i=\frac{h_m+h_l}{1+h_l}$ if $i = m$ and $\frac{hi}{1+h_l}$, otherwise, and c) $\mu \leq 1+h_l$. In this case, the payoff of the attacker remains constant.
\end{theorem}
\vspace*{-0.3cm}
\begin{proof}\let\qed\relax
	If $\mu \leq 1+h_l$, then $\Omega _A=\emptyset$ and (\ref{eq:main}) will become:
	\begin{equation*}
	\begin{aligned}
	\mu ^* _1 \sum_{i\notin \Omega _A}h_i-\frac{R_D}{R_A}\sum_{i\notin \Omega _A} \frac{\left(h_i\right)^2}{g_i}=0,\\
	\mu ^* _1-\frac{R_D}{R_A}\left(\sum_{i=1 i\neq m}^{n}\left(\frac{\left(h_i\right)^2}{\frac{h_i}{1+h_l}}\right)+\frac{\left(h_m\right)^2}{\frac{h_m+h_l}{1+h_l}}\right)=0,\\
	\mu ^* _1=\frac{R_D}{R_A}\left(\left(1+h_l\right)\left(1-h_m\right)+\left(1+h_l\right)\left(\frac{\left(h_m\right)^2}{h_m+h_l}\right)\right),\\
	\end{aligned}
	\end{equation*}
	\begin{equation}\label{eq:mu1}
	\mu ^* _1=\frac{R_D}{R_A}\left(\left(1+h_l\right)\left(\frac{h_m+h_l-h_mh_l}{h_m+h_l}\right)\right).
	\end{equation}
	Now, to check the condition of above solution we have:
	\begin{equation*}
	\begin{aligned}
	\mu ^* _1 &\ge 1+h_l, \\
	\frac{R_D}{R_A}\left(\frac{h_m+h_l-h_mh_l}{h_m+h_l}\right)&\ge 1,\\
	\end{aligned}
	\end{equation*}
	\begin{equation}\label{maxmufirst}
	\frac{R_D}{R_A}\ge \frac{h_m+h_l}{h_m+h_l-h_mh_l}.
	\end{equation}
	\eqref{maxmufirst} is the condition to have the $\mu ^*_1$ as a valid solution for  (\ref{eq:main}). If $\mu ^* _1$ satisfies its condition, then $\lambda _A$ and $\lambda _D$ are obtainable as:
	\begin{equation}\label{eq:lambdaA}
	\lambda _A=\frac{1}{2R_D}, \lambda _D=\frac{1}{2R_A} \frac{h_m+h_l}{\left(1+h_l\right)\left(h_m+h_l-h_mh_l\right)}.
	\end{equation}
	Then, the expected payoffs for defender and attacker at Nash equilibrium are:
	\begin{equation}\label{eq:pi}
	\pi _A=\frac{R_A}{2R_D}, \pi _D=1+\frac{R_D-2R_A}{2R_D}\frac{h_m+h_l}{\left(1+h_l\right)\left(h_m+h_l-h_mh_l\right)}.
	\end{equation}
	Also, when the attacker has complete information, one can easily show that the  payoffs for both players are:
	\begin{equation}\label{eq:pip}
	\pi _A=\frac{R_A}{2R_D}, \pi _D=1-\frac{R_A}{2R_D}.
	\end{equation}
\end{proof}
By comparing \eqref{eq:pip} with \eqref{eq:pi}, we can see that the defender gains more in the case of no information for the attacker, however, the attacker's payoff remains constant.
\subsection{Numerical Case Study}\vspace*{+0.05cm}
Next, in addition to the mathematical analysis, we will define human interaction, capacity and flows of interconnections, and minimum path parameters for a CPS and we will numerically analyze the effect of $v_{ij}$ as given by \eqref{eq:6}. For our simulations, we consider a CPS with 9 cyber nodes controlling 9 physical nodes as in Figure \ref{fig:example}. Also, we consider three levels of values for the nodes. The reference node will have the highest value, followed by the main nodes, and finally the ordinary nodes. As mentioned earlier, we assume that $h_i$ depends on the distance from the reference node and it is also equal for all the nodes at the same level. Therefore, in Figure \ref{fig:example}, $h_1=3h_4=5h_7$. The second column of Table \ref{tab:HI} shows the human interactions of each node. All simulations pertain to the case in which $\frac{R_D}{R_A}=2.5$.

In Table I, we consider a system in which the interdependencies lead to an increase in the value of only one node at each level while decreasing the values of all other CPS nodes. Here, we consider three cases, shown in Table \ref{tab:HI}. In each case we just increase the value of only one node in one level and decrease all the other values. The payoffs in Table \ref{tab:HI} show that, in case 3, the defender's payoff increases more than in the other cases due to the more symmetric values for $g_i$.

\begin{figure}[!t]
	\centering
	\includegraphics[width=7cm]{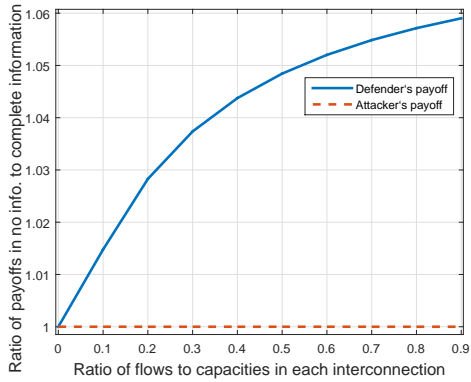}
	\caption{Ratio between the payoffs in the game with no information and the game with complete information, as the ratio between flows and capacities varies.}
	\label{fig:ratiocapcaity}
	\vspace*{-0.15cm}
\end{figure}

Next, we analyze the system by increasing the flows of interconnections. Figure \ref{fig:ratiocapcaity} shows that, as the ratio of flow to capacity at each interconnection increases, the payoff for the defender increases, however, the payoff for the attacker remains constant. As the ratio of flows to capacities increases, all the values for $g_i$ become less asymmetric which, for our CPS, correspond to a value of $\frac{1}{9}$.

\begin{figure}[!t]\vspace*{-0.2325cm}
	\centering
	\includegraphics[width=7cm]{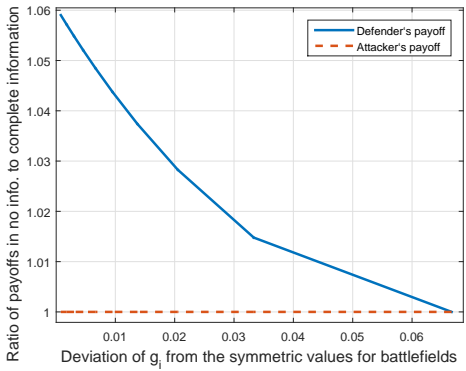}
	\caption{Ratio of player's payoffs in the incomplete information case to the case with complete information, as the nodes become less asymmetric.}
	\label{fig:ratiovariance}
	\vspace*{-0.25cm}
\end{figure}

Figure \ref{fig:ratiovariance} shows that as the nodes become more symmetric, the payoff of the defender increases. However, the attacker's payoff remains constant. From Figure \ref{fig:ratiovariance}, we can see that, as the nodes become more symmetric, the defender's payoff increases up to 6\% compared to the case in which the system has more.
\begin{table}[!t]
	%\scriptsize
	\centering
	\caption{%\mycaption{%\vspace*{-1em}
		\vspace*{-0.4em}Human Interaction of interdependent and independent nodes}\vspace*{-1em}
	\begin{tabular}{ |p{1.5cm}|p{1cm}|p{1.5cm}|p{1.5cm}|p{1.5cm}|}
		\hline
		\hspace*{1em}$i$ & \hspace*{1em}$h_{i}$ & $g_i$ for case 1& $g_i$ for case 2& $g_i$ for case 3\\
		\hline
		\hspace*{1em}1&$0.2667$&\hspace*{1em}$0.3282$&\hspace*{1em}$0.2406$&\hspace*{1em}$0.2388$\\
		\hline
		\hspace*{1em}2&$0.1333$&\hspace*{1em}$0.1221$&\hspace*{1em}$0.2180$&\hspace*{1em}$0.1194$\\
		\hline
		\hspace*{1em}3&$0.1333$&\hspace*{1em}$0.1221$&\hspace*{1em}$0.1203$&\hspace*{1em}$0.1194$\\
		\hline
		\hspace*{1em}4&$0.1333$&\hspace*{1em}$0.1221$&\hspace*{1em}$0.1203$&\hspace*{1em}$0.1194$\\
		\hline
		\hspace*{1em}5&$0.0667$&\hspace*{1em}$0.0611$&\hspace*{1em}$0.0602$&\hspace*{1em}$0.0597$\\
		\hline
		\hspace*{1em}6&$0.0667$&\hspace*{1em}$0.0611$&\hspace*{1em}$0.0602$&\hspace*{1em}$0.0597$\\
		\hline
		\hspace*{1em}7&$0.0667$&\hspace*{1em}$0.0611$&\hspace*{1em}$0.0602$&\hspace*{1em}$0.0597$\\
		\hline
		\hspace*{1em}8&$0.0667$&\hspace*{1em}$0.0611$&\hspace*{1em}$0.0602$&\hspace*{1em}$0.0597$\\
		\hline
		\hspace*{1em}9&$0.0667$&\hspace*{1em}$0.0611$&\hspace*{1em}$0.0602$&\hspace*{1em}$0.1641$\\
		\hline
		Def. Payoff&\hspace*{1em}$0.8$&\hspace*{1em}$0.8034$&\hspace*{1em}$0.8081$&\hspace*{1em}$0.8130$\\
		\hline
		At. Payoff&\hspace*{1em}$0.2$&\hspace*{2em}$0.2$&\hspace*{2em}$0.2$&\hspace*{2em}$0.2$\\
		\hline
	\end{tabular}\vspace*{-0.48cm}\label{tab:HI}
\end{table}

\begin{figure}[!t]
	\centering
	\includegraphics[width=7.6cm]{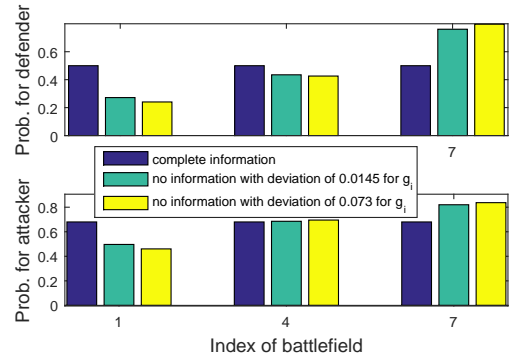}
	\caption{Probability of allocating $g_i$ fraction of resources for defender and $h_i$ fraction of resources for attacker in three levels of nodes.}
	\label{fig:prob}
	\vspace*{-0.25cm}
\end{figure}

Furthermore, to analyze how the interdependencies impact the allocation of resources, we calculate the probability of allocating resources over three nodes from three different levels of human interaction proportional to their value. Figure \ref{fig:prob} shows that as nodes become more symmetric, both attacker and defender tend to allocate resources proportional to their value with higher probability to the node with lower human interaction. Note that, a decrease in deviation of $g_i$ captures the increase in the interdependencies.
\section{Conclusions}\label{sec:Conclusion}
In this paper, we have proposed a novel framework for analyzing the security of a CPS with interdependent cyber and physical nodes. In particular, we have modeled the interdependencies using notions of flow capacity and minimum path. Then, for the modeled system, we have formulated a novel Colonel Blotto game in which an attacker seeks to compromise the CPS by allocating its destructive resources on cyber nodes, and a defender aims to protect the CPS by allocating defensive resources. For this game, we have analyzed the equilibrium strategies, and we have shown various properties for the particular case in which interdependencies only increase the value of reference node. Simulation results have also shown that, for the general game, the defender can increase its payoff in the case of high interdependency and no information for attacker.

\bibliographystyle{ACM-Reference-Format}
\bibliography{references} 

\end{document}